\documentclass[aps,showpacs,preprint]{revtex4}
\usepackage{graphicx}
\begin{document}
\newcommand{\be}{\begin{equation}}
\newcommand{\ee}{\end{equation}}
\newcommand{\bea}{\begin{eqnarray}}
\newcommand{\eea}{\end{eqnarray}}
\title{Folding Mechanism of Small Proteins}
\author{Seung-Yeon Kim$^1$, Julian Lee$^{1,2}$,
and Jooyoung Lee$^1$\footnote{correspondence to: jlee@kias.re.kr}}
\affiliation{$^1$School of Computational Sciences, Korea Institute for Advanced Study,
Seoul 130-012, Korea   \\
$^2$ Department of Bioinformatics and Life Sciences, Soongsil
University, Seoul 156-743, Korea }
\begin{abstract}
Extensive Monte Carlo {\it folding} simulations for four proteins
of various structural classes are carried out,
using a {\it single} atomistic potential.
In all cases, collapse occurs
at a very early stage, and proteins fold into their native-like conformations
at appropriate temperatures. The results demonstrate that
the folding mechanism is controlled not only by thermodynamic factors
but also by kinetic factors:
The way a protein folds into its native structure, is also determined
by the convergence point of early folding trajectories,
which cannot be obtained by the free energy surface.
\end{abstract}
\pacs{87.14.Ee, 87.15.$-$v, 05.20.Dd, 05.70.$-$a}

\maketitle
Understanding how a protein folds is a long-standing challenge in
modern science. Computer simulations
\cite{duan,skolnick,dinner,go,zhou,galzitskaya,shea,kuss,
alonso,boczko,bursulaya,fernandez,ghosh}
have been carried out to understand the folding mechanism. However,
simulation of protein-folding processes with an atomistic model is
a very difficult task. The difficulties come from two sources: (i)
Direct folding simulation using an all-atom potential requires
astronomical amounts of CPU time, and typical simulation times are
only about a few nanoseconds. (ii) Atomic pairwise interactions
including solvation effects may not be accurate enough. An
extensive folding simulation has been carried out for the villin
headpiece subdomain (HP-36), where 1-$\mu s$ molecular dynamics
simulation with an all-atom potential has been performed,
producing only candidates for folding intermediates \cite{duan}. For this
reason, direct folding simulations have been mainly focused on
simple models, such as lattice models \cite{skolnick,dinner}, models where only
native interactions are included (Go-type models) \cite{go,zhou,galzitskaya,shea},
and a model with discrete energy terms whose parameters are optimized
{\it separately} for each protein \cite{kuss}.
Alternative indirect approaches have also been proposed including
unfolding simulations \cite{shea,alonso,boczko,bursulaya}
starting from the folded state
of a protein. However, it is not obvious whether the folding is
the reverse of the unfolding \cite{dinner,shea}. Moreover, to the best of our knowledge,
no one has yet succeeded in
folding more than one proteins into their native states
using a {\it single} potential,
even with simplified models \cite{kuss}.

In this letter, we propose a novel method to fold several of small proteins
using Monte Carlo dynamics.
This method uses a {\it single} atomistic continuous potential
which includes {\it all pairwise} (native and non-native) interactions,
and yet allows us
to carry out {\it folding} simulations starting from non-native
conformations.
We observe that all proteins fold into their native-like
conformations at appropriate temperatures. We find that the
folding mechanism is controlled by both kinetic and thermodynamic
factors \cite{lee1}: The way a protein folds into its native structure, is
determined not only by the free energy surface,
but also by the convergence point of early folding trajectories
relative to the native state.

We consider a system where an off-lattice
potential energy function including non-native interactions is
utilized. We study the folding dynamics of proteins using a
united-residue (UNRES) \cite{liwo} force field where a protein is
represented by a sequence of $\alpha$-carbons linked by virtual
bonds with attached united side chains and united peptide groups.
Energy terms are all continuous, and include pairwise electrostatic,
van der Waals, and multibody terms (see Ref. \cite{liwo} for details).
The effect of solvent was indirectly included in the force field.
The parameters of the UNRES force field were {\it simultaneously}
optimized \cite{lee2} for four proteins of betanova \cite{bursulaya,kortemme} (20
residues, three-stranded $\beta$-sheet), 1fsd \cite{dahiyat} (28 residues,
one $\beta$-hairpin and one $\alpha$-helix), HP-36 \cite{duan,fernandez} (36
residues, three-helix bundle), and fragment B of staphylococcal
protein A \cite{zhou,shea,alonso,boczko,ghosh,bai} (46 residues, three-helix
bundle). The low-lying local-energy minima for these proteins were
found by the conformational space annealing \cite{lee3} method. The
parameters were modified in such a way that the native-like
conformations are energetically more favored than the others. The
global minimum-energy conformations found using the optimized
force field are of the root-mean-square deviation (RMSD) values
1.5 {\AA}, 1.7 {\AA}, 1.7 {\AA} and 1.9 {\AA} from the {\it experimental}
structures for
betanova, 1fsd, HP-36 and protein A, respectively. After the
parameter optimization, {\it one} set of the parameters is obtained for
four proteins.

In the UNRES force field there are two backbone angles and two
side chain angles per residue (no side-chains for glycines). The
values of these angles are perturbed one at a time, typically
about $15^\circ$, and the backbone angles are chosen three times more
frequently than the side chain angles. The perturbed conformation
is accepted according to the change in the potential energy,
following Metropolis rule. Since only small angle changes are
allowed one at a time, the resulting Monte Carlo dynamics can be
viewed as equivalent to the real dynamics. At a fixed temperature,
at least ten independent simulations starting from various
non-native states of a protein were performed up to $10^9$ Monte
Carlo steps (MCS) for each run. During simulation the values of
RMSD from the native structure and the radius of gyration ($R_g$) were calculated
using $C_\alpha$ coordinates. The lowest RMSD values from folding simulations are
0.78 {\AA}, 1.07 {\AA}, 1.58 {\AA} and 2.07 {\AA} for betanova, 1fsd, HP-36 and
protein A, respectively. The fractions of the native contacts ($Q$
and $\rho$) were also measured during simulations, where $Q$ is
calculated from the experimental structure. A native contact is
defined to exist when two $C_\alpha$'s separated more than two residues in
sequence are placed within 7.0 {\AA}. To define $\rho$ we first
characterize the native state conformations by performing
simulations starting from the experimental structures, at the same
temperatures where folding simulations were performed. We define $\rho$
as the fraction of native contacts weighted over their contact
probabilities in the native state simulations \cite{shea,boczko,bursulaya}. Due to
the fluctuation of the native conformation, the value of $Q$ is
usually lower than that of $\rho$

The time histories of the typical trajectories from the folding
simulations of betanova and HP-36 are shown in Fig.~1. The
trajectories for 1fsd and protein A are similar to those in the
figure. We observe that collapse occurs at a very early stage
($\sim10^4$ MCS) for all four proteins, but the details of each folding
process appear to depend on the secondary structure contents.
Distributions of RMSD, $Q$, $\rho$ and $R_g$ are also accumulated during the
whole simulations. The distributions of RMSD are shown in Fig.~2,
and those for $\rho$ and $R_g$ are shown in Fig.~3 as contour plots. The
early folding trajectories plotted in Fig.~3 are obtained as
follows. We divided the initial $10^5$ MCS into 19 intervals (ten $10^3$
MCS and subsequent nine $10^4$ MCS), and took average over
conformations in each interval. These averages were again averaged
over 100 independent simulations starting from random
conformations. The same procedure was applied to 100 independent
simulations starting from a fully extended conformation.

The simulation of betanova at $T=40$ (arbitrary unit) starting from
an unfolded conformation demonstrates that rapid collapse occurs
in about $10^4$ MCS; the value of $R_g$ decreases, whereas the value of
$Q$ remains below 0.3 (Fig.~1). During the next $10^8$ MCS, more compact
states appear, and the value of $Q$ moves up as high as 1.0. There
are two values of RMSD that are populated (see also Fig.~2(b)): one
corresponds to less ordered structures (RMSD$\sim$5 {\AA}) and the other to
native-like structures (RMSD$\sim$2.5 {\AA}). This demonstrates, although
the native-like structure is the most stable one, thermal
fluctuation can temporarily kick the protein out of the native
structure.

We now analyze the details of the folding behavior of each
protein. For betanova at low temperatures ($T\le30$), the probability
distributions of various quantities such as RMSD depend on initial
conformations, showing its glassy behavior (Fig.~2(a)). At higher
temperatures ($T\ge40$) this non-ergodic glassy behavior disappears.
It should be noted that native-like structures are more easily
found from the simulation at $T=40$ (Fig.~2(b)) than from the best of
ten runs at $T=30$. When temperature decreases from $T=80$ to 60, the
location of the RMSD peak dramatically moves from 8 {\AA} to 3 {\AA}. This
demonstrates the cooperative folding characteristics of betanova.
For $40\le T\le60$ betanova folds into its native-like structure. The
initial folding trajectories and the distribution of ($\rho$, $R_g$) at
$T=40$ are shown in Fig.~3(a) for betanova. Regardless of its initial
conformation (either random or fully extended), the average
pathways to the folded conformation initially converge to
($\rho$, $R_g$)$\sim$(0.35, 8.5 {\AA}), and then they move horizontally
to the native structure. This is consistent with the recent folding scenario for
proteins with $\beta$ structure \cite{shea}.
The populated states \cite{bursulaya} around
$\rho \sim $0.4-0.5 and $R_g\sim$11-12 {\AA}
are not from initial folding trajectories,
but from the fluctuation of native-like structures. This kinetic
information is difficult to be captured by free energy
calculations alone \cite{bursulaya}.

For 1fsd the distributions of various quantities for ten
independent runs ($10^9$ MCS each) show glassy behavior for $T\le50$.
Again, the non-ergodic glassy behavior disappears at higher
temperatures ($T\ge70$). Fig.~2(c) shows the RMSD distributions at
various temperatures. The strength of the cooperativity for 1fsd
is not as strong as that of betanova. Again after initial collapse
to ($\rho$, $R_g$)$\sim$(0.3, 9 {\AA}), the average trajectories move horizontally
(Fig.~3(b)), although less prominently so compared to the case of
betanova.

For HP-36, the distributions of various quantities for ten
independent runs ($10^9$ MCS each) again show glassy behavior at
$T=60$. At higher temperatures we have two RMSD peaks (Fig.~2(d)). At
$T=90$ the peak near the native structure begins to dominate, and as
temperature decreases it becomes stronger. This double peak
feature demonstrates the cooperative two-state transition. The
conformations centered at the higher value of RMSD come from a
variety of collapsed states. The conformations from the other peak
are native-like. When we examine them, the helix I (residues 4-8)
is stably formed \cite{fernandez}, while the others are fluctuating. The
initial folding trajectories (Fig.~3(c)) converge to ($\rho$, $R_g$)$\sim$(0.2,
11 {\AA}). These collapsed structures fold into native-like structures
($\rho$, $R_g$)$\sim$(0.7, 10 {\AA}). Compared to the case of betanova and 1fsd, the
average folding trajectories are more diagonal.

The overall folding characteristics of protein A are similar to
those of HP-36. The initial folding trajectories (Fig.~3(d))
converge to ($\rho$, $R_g$)$\sim$(0.25, 12 {\AA}). These collapsed states fold into
native-like structures in a diagonal fashion similar to the case
of HP-36. When we examine the native-like conformations with
$3 {\rm{\AA}}\le {\rm RMSD}\le4 {\rm{\AA}}$,
the helix III (residues 42-55) is most stably formed.
This is in agreement with recent investigations \cite{alonso,ghosh,bai}. We
also observe that a first-order like collapse transition \cite{zhou}
(from ($Q$, $R_g$)$\sim$(0.15, 18 {\AA}) to (0.15, 12 {\AA}))
occurs near $T=120$.

By using an atomistic model, we could observe folding processes of
four small proteins in realistic settings. In all cases, rapid
collapse is followed by subsequent folding process that takes
place in a longer time scale. The folding mechanism suggested in
this study is as follows: There are two aspects of folding
dynamics, (i) non-equilibrium kinetic properties and (ii)
equilibrium thermodynamic properties (Fig.~4). The non-equilibrium
kinetic properties, relevant to the early folding trajectories
(fast process), can be examined only by direct folding
simulations. The free energy surface, an equilibrium thermodynamic property, dictates the
way an initially collapsed state completes its folding (slow
process). The way a protein folds into its native structure, i.e.,
either horizontally or diagonally in ($\rho$, $R_g$) plane, is determined
by the position of ($\rho$, $R_g$) where early folding trajectories
converge, relative to the native state. It appears that slow
folding process of $\alpha$-proteins occurs in a diagonal fashion
compared to that of proteins containing $\beta$-strands \cite{shea}.

In conclusion, we successfully carried out direct folding simulations of
more than one proteins using a {\it single} atomistic potential.
We also observe that glassy transitions occur at low temperatures.
The results provide new insights into the folding
mechanism.

\newpage

\begin{figure}
\includegraphics[angle=270,width=15cm]{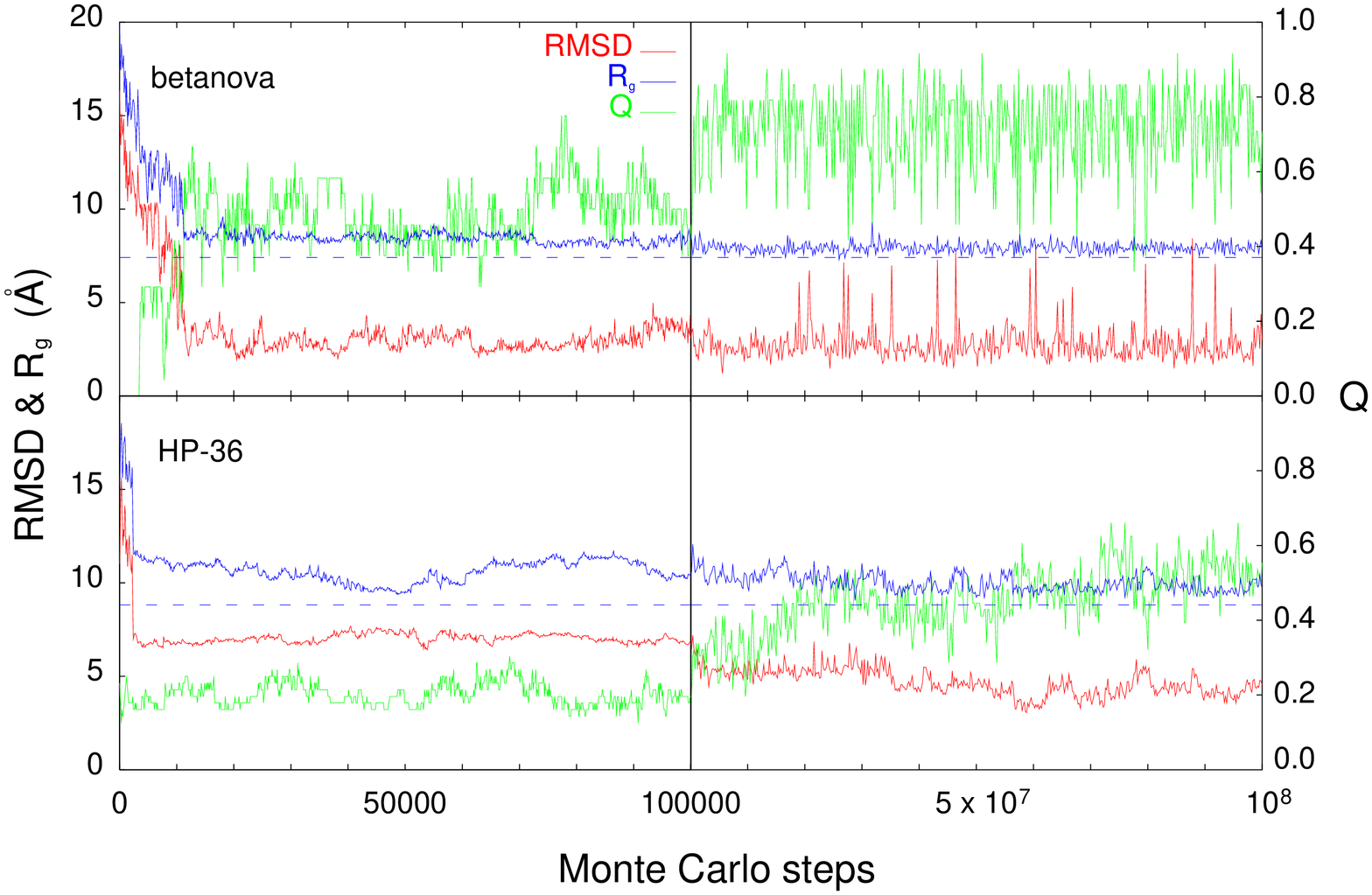}
\caption{Typical folding trajectories of
betanova at $T=40$ and HP-36 at $T=70$ starting from non-native
conformations. RMSD, $R_g$ and $Q$
are plotted for every 100 MCS in the early part
and for every $2 \times 10^5$ steps in the subsequent
part. The dotted lines represent the values of $R_g$ of the native
states.}
\end{figure}

\begin{figure}
\includegraphics[angle=270,width=7cm]{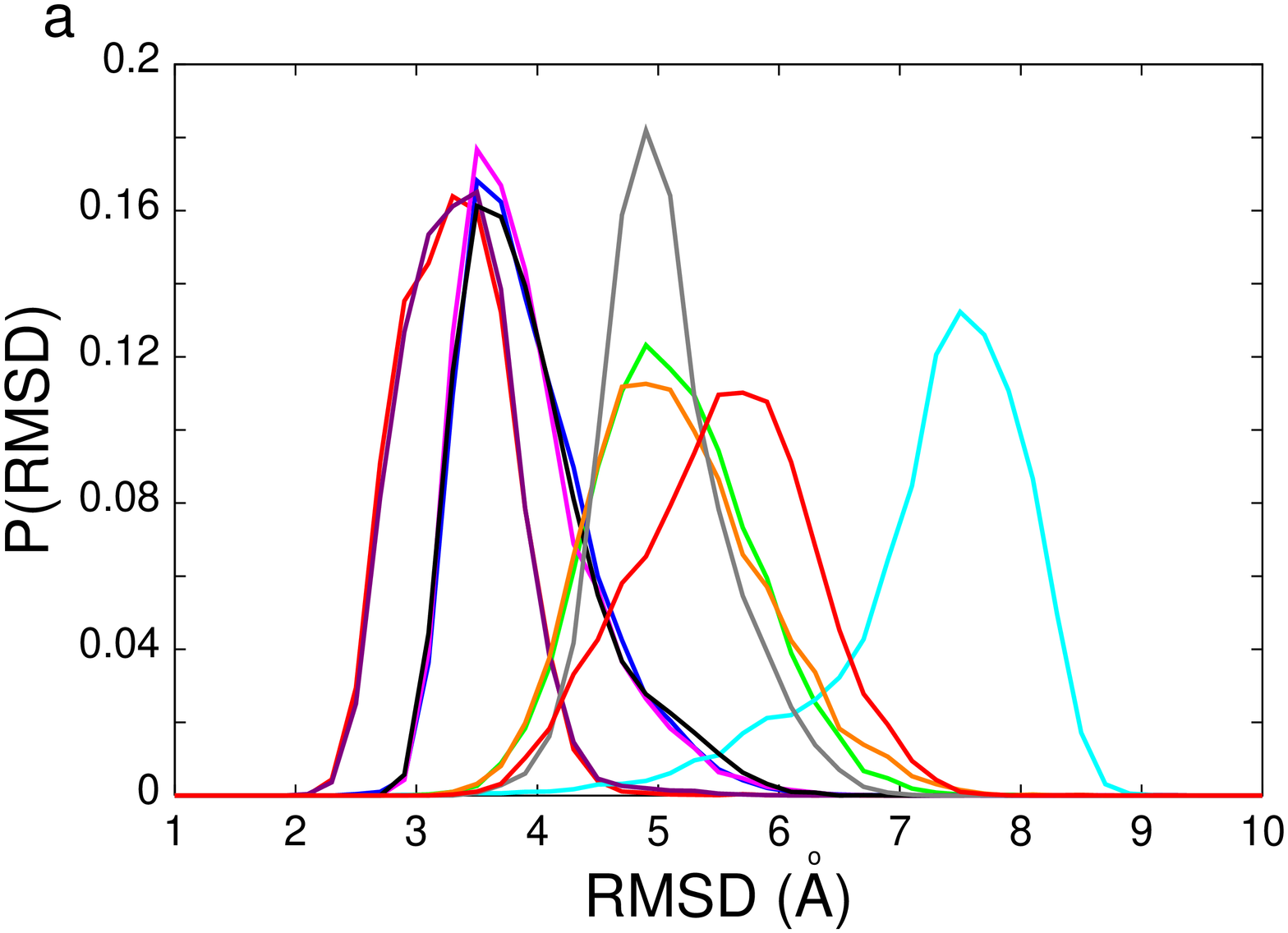}
\includegraphics[angle=270,width=7cm]{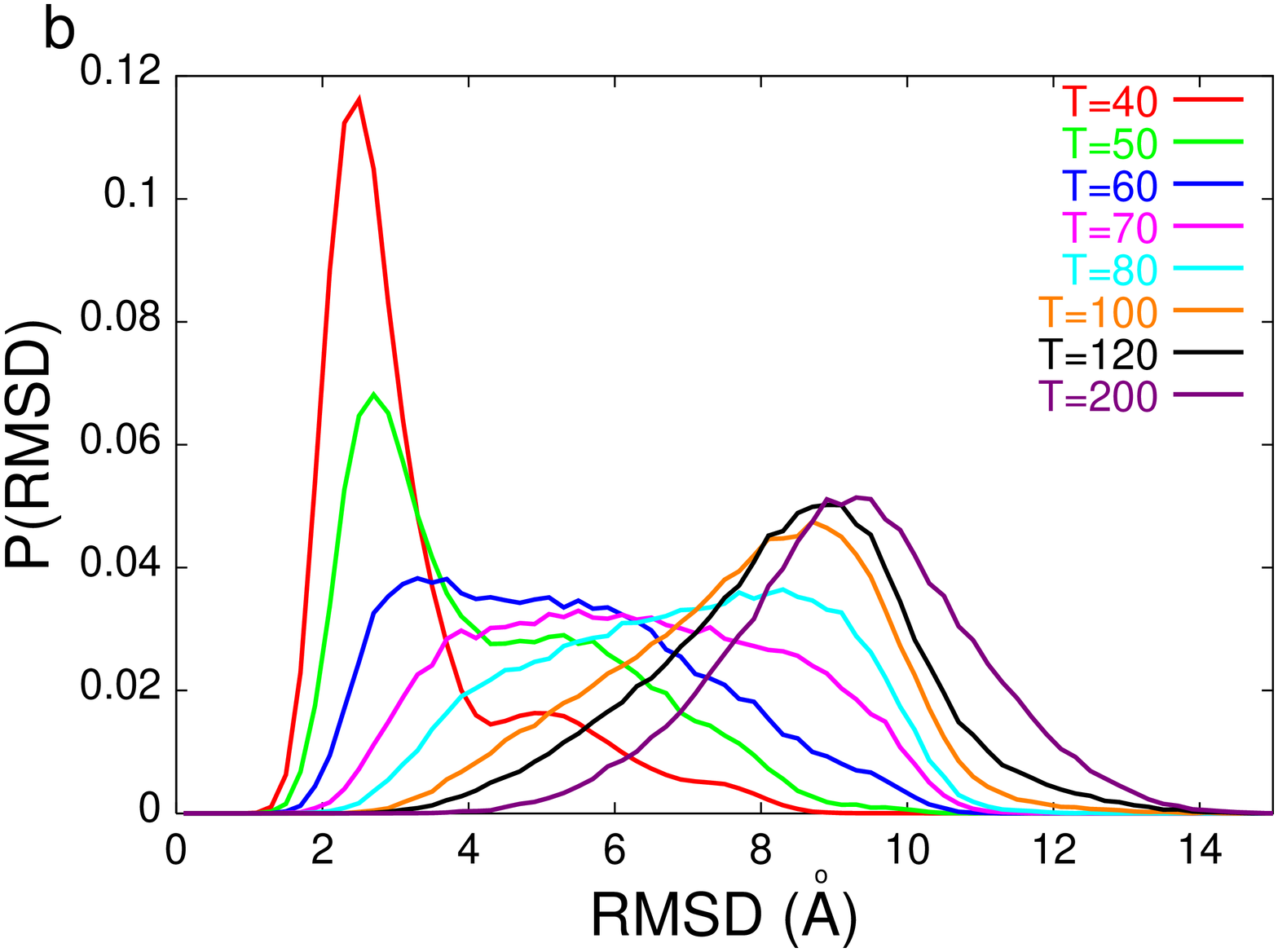}
\includegraphics[angle=270,width=7cm]{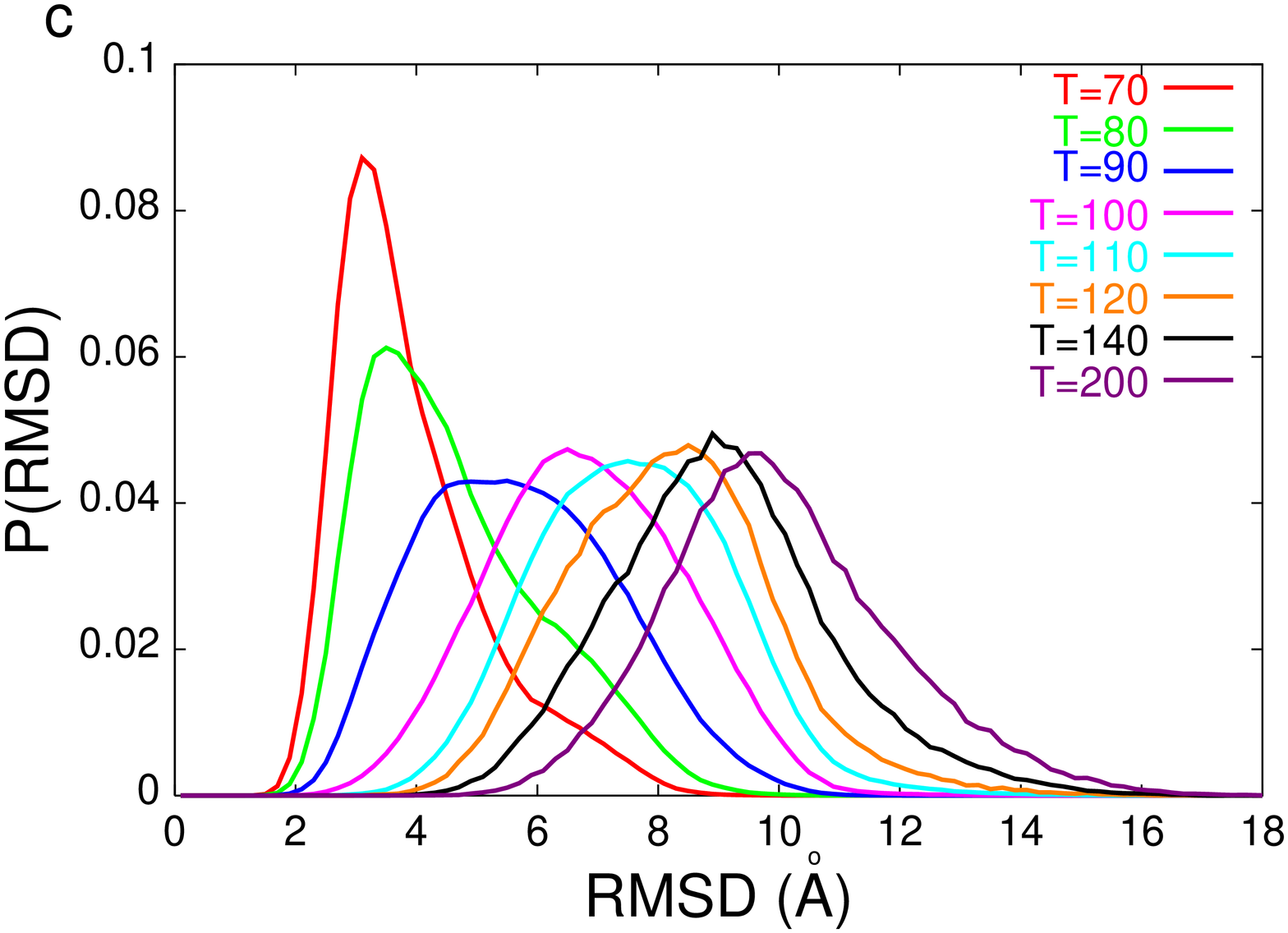}
\includegraphics[angle=270,width=7cm]{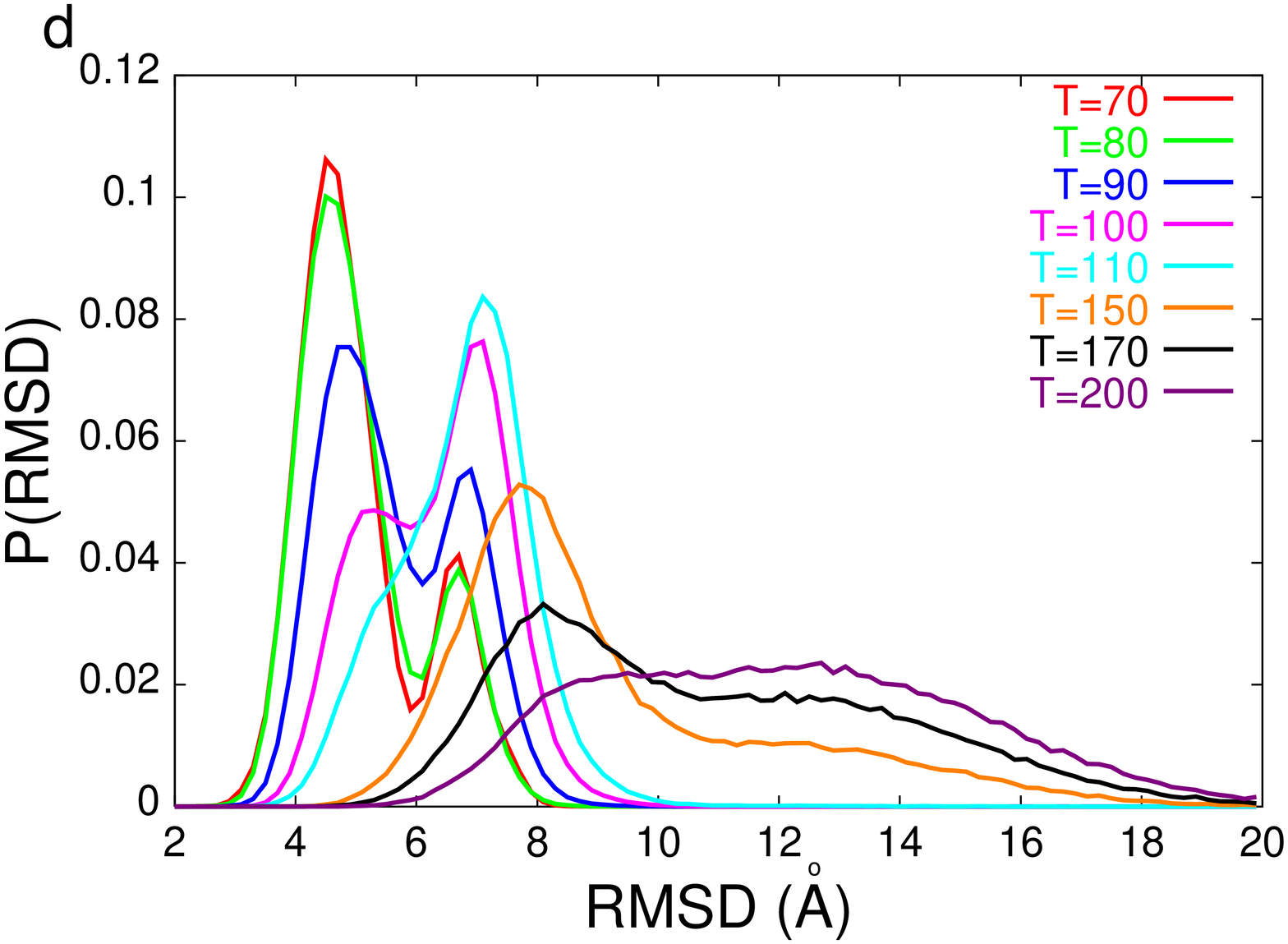}
\caption{The probability distributions of RMSD. (a) The
distributions of 10 simulations for betanova, starting from random
conformations at $T=30$.
The glassy behavior is apparent. (b) RMSD
distributions of betanova at various temperatures. As temperature
decreases, the value of prominent RMSD dramatically moves from
9.0 {\AA} to 2.5 {\AA}. (c) The distributions for 1fsd. The value of the most
probable RMSD drops rapidly from 9.0 {\AA} to 3.1 {\AA}. (d) The
distributions for HP-36. The double peak structure appears for
$T\le100$, representing the cooperative (first-order like) two-state transition.}
\end{figure}

\begin{figure}
\includegraphics[width=7cm]{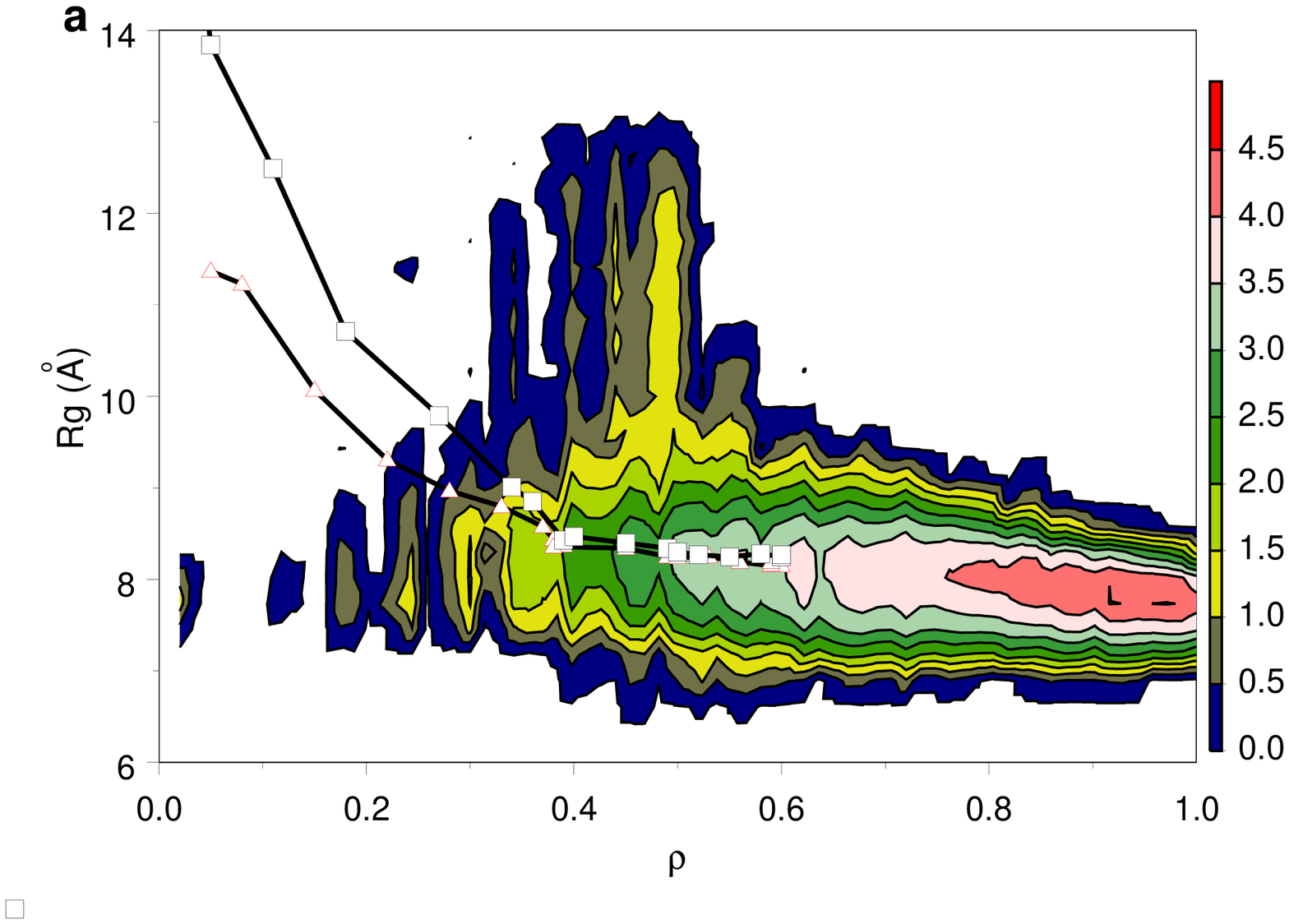}
\includegraphics[width=7cm]{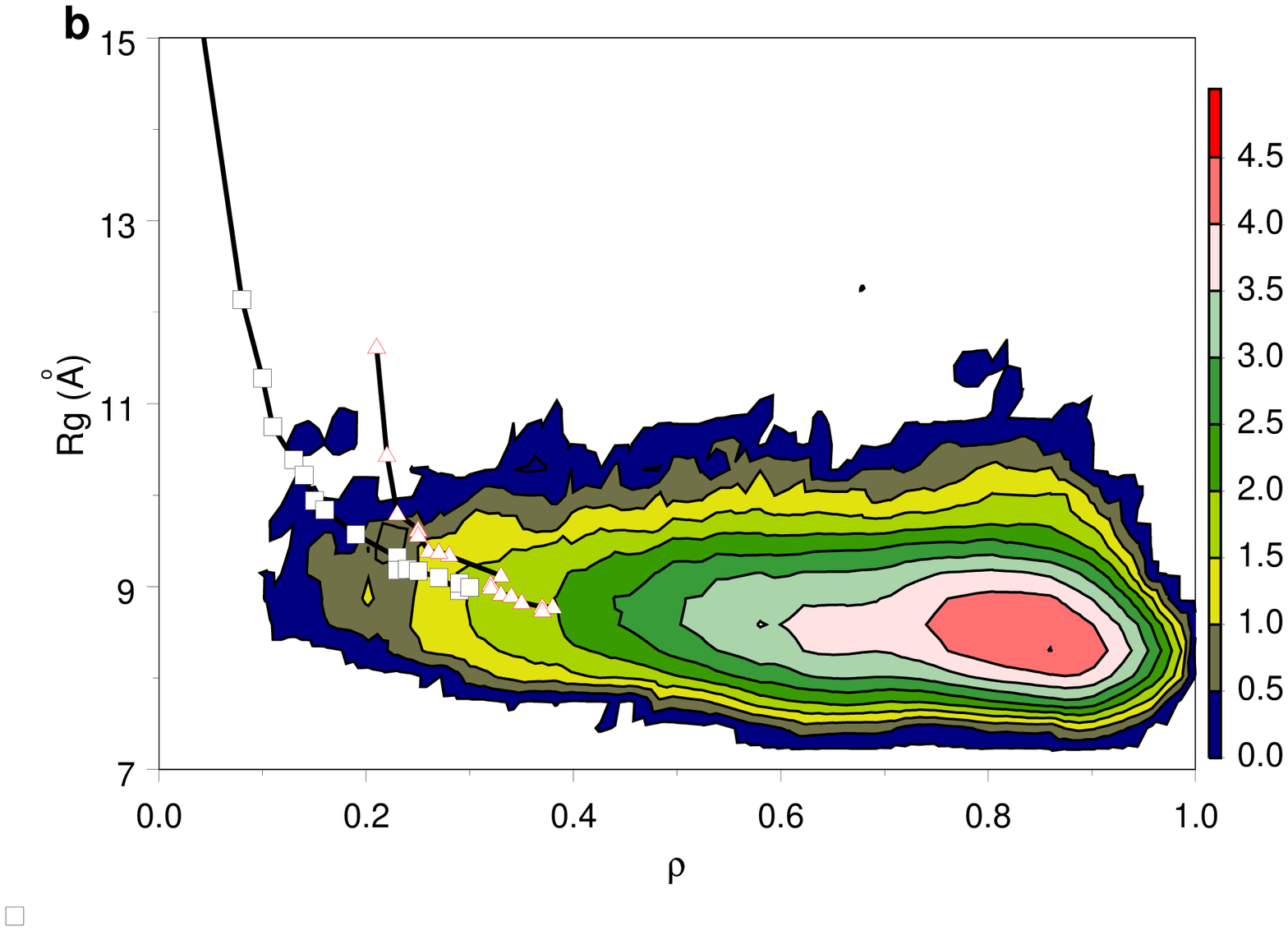}
\includegraphics[width=7cm]{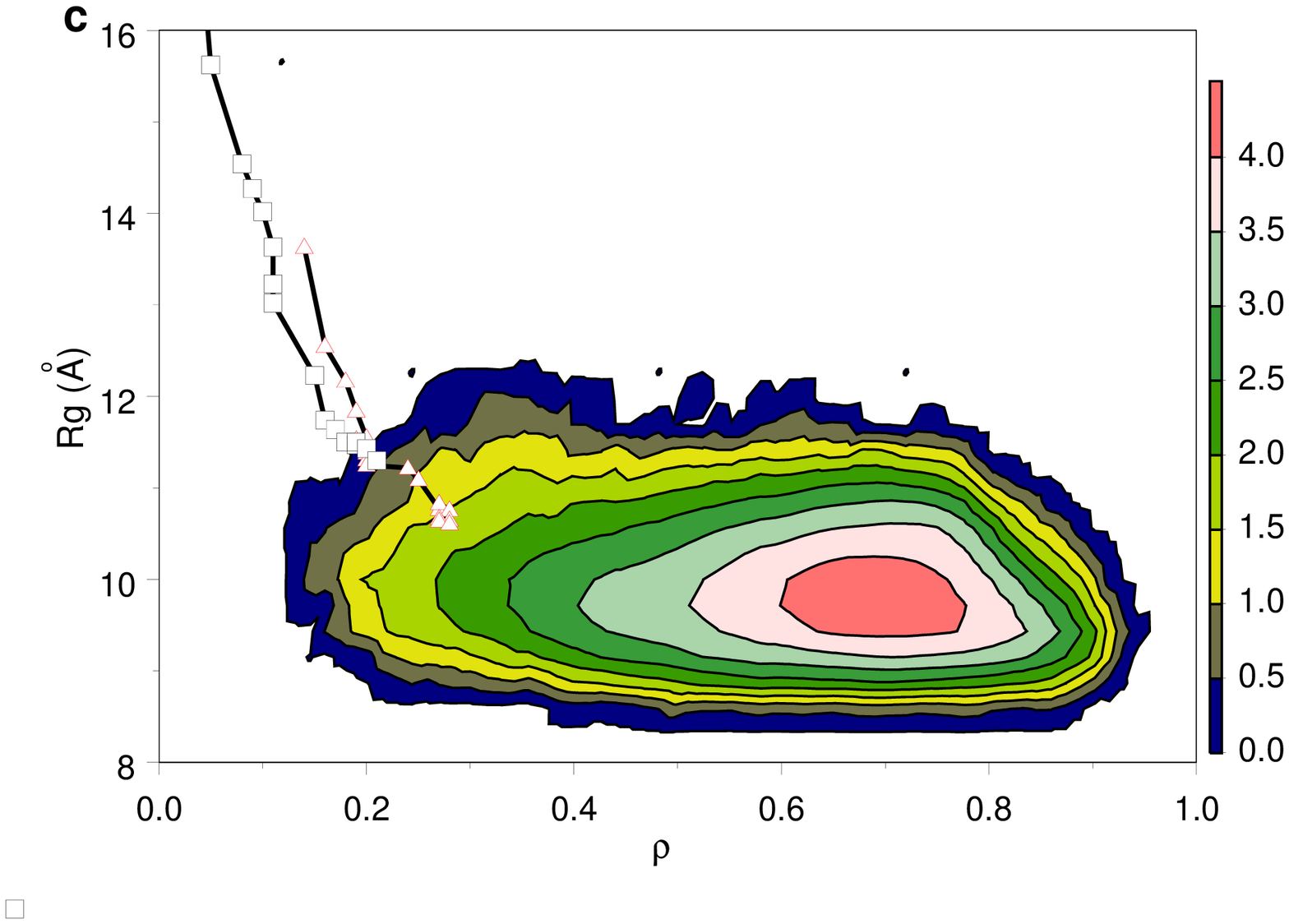}
\includegraphics[width=7cm]{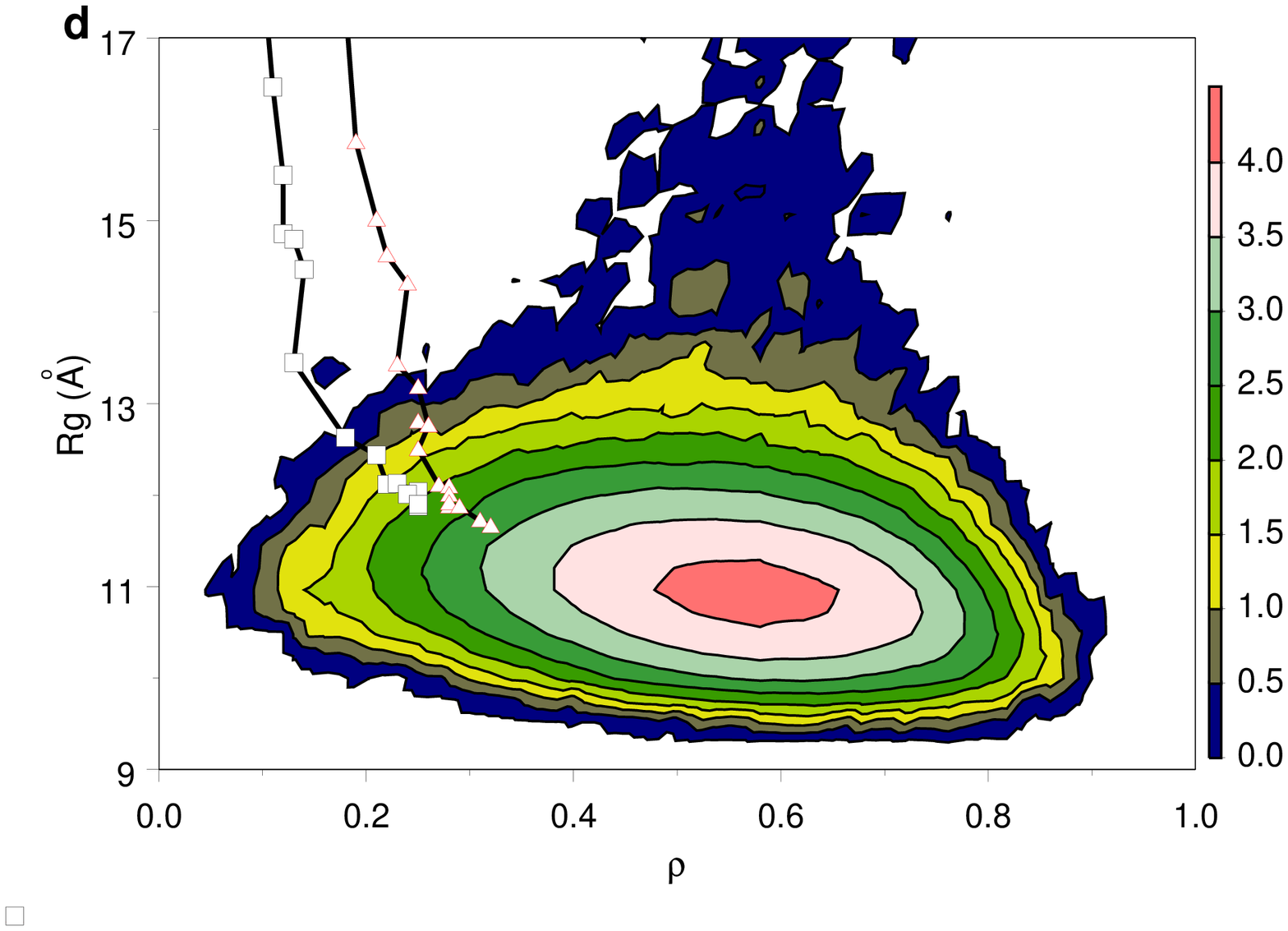}
\caption{The initial folding trajectories and the contour plots of
the population distributions as a function of $\rho$
and $R_g$ at appropriate temperatures.
The triangles represent the averages of 100 folding trajectories
starting from random conformations. The squares are from
100 trajectories starting from a fully extended conformation.
The color scale indicates the exponent $x$
of the population $10^x$ at given values of $\rho$ and $R_g$. (a) betanova at
$T=40$. (b) 1fsd at $T=70$.
(c) HP-36 at $T=70$. (d) protein A at $T=80$.}
\end{figure}

\begin{figure}
\includegraphics[width=15cm]{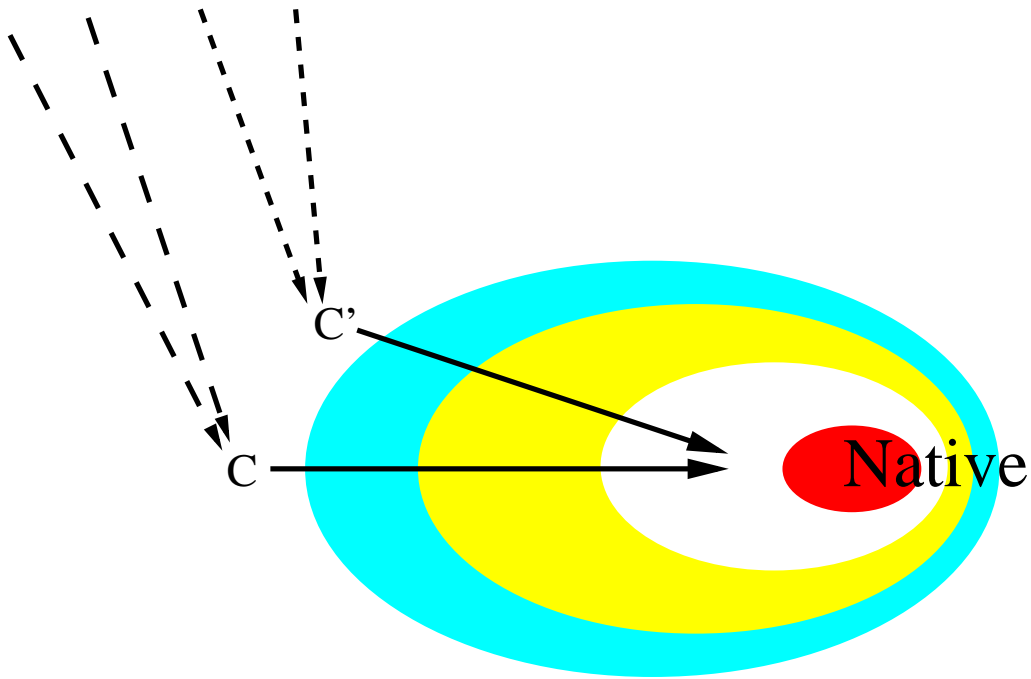}
\caption{A schematic of the folding trajectories. The
colored contour represents the free energy surface, which is
an equilibrium property. Even for proteins with identical free
energy landscape, the early folding trajectories (dashed lines)
may converge into different points (C or C'). The solid lines
represent the later part of the folding trajectories dictated by
the free energy landscape. The position of the convergence point
of a protein is determined by its kinetic properties. This
information can be obtained only by direct {\it folding} simulations.}
\end{figure}

\end{document}